\documentclass{aa}

\newcommand{\rs}{R$_{\odot} $\,}

\begin{document}

\title{ Near-Sun and 1 AU magnetic field 
of coronal mass ejections: A parametric study}

\author{S. Patsourakos\inst{1} \and M. K.  Georgoulis\inst{2}}

\offprints{S. Patsourakos, \email{spatsour@cc.uoi.gr}}

\institute{University of Ioannina, Department of Physics,
Section of Astrogeophysics, Ioannina, Greece.  \and
Research Center for Astronomy and Applied Mathematics, Academy of Athens, Athens, Greece.}

\abstract
{}
{The  magnetic field of coronal mass ejections (CMEs) determines their structure, evolution, and  energetics, 
as well as  their geoeffectiveness. However, we currently lack routine diagnostics of the near-Sun CME magnetic field, which is crucial for determining the subsequent evolution of CMEs.}
{We recently presented a method to infer the near-Sun magnetic field magnitude of CMEs and then extrapolate it to 1 AU. 
This method uses relatively easy to deduce observational estimates of the magnetic helicity in CME-source regions 
along with geometrical CME fits enabled by coronagraph observations. 
We hereby perform a parametric study of this method
aiming to assess its robustness.
We use statistics of active region (AR) helicities and CME geometrical parameters 
to determine a matrix of plausible near-Sun CME magnetic field magnitudes. In addition, 
we extrapolate this matrix to 1 AU and determine the anticipated range 
of CME magnetic fields 
at 1 AU representing the radial falloff of the magnetic
field in the CME out to interplanetary (IP) space by a power law with index
${\alpha}_{B}$.}
{The resulting distribution of the near-Sun (at 10 \rs)
CME magnetic fields varies in the range  [0.004, 0.02] G, comparable to, or higher than, a few existing observational
inferences of the magnetic field in the quiescent corona
at the same distance. We also find that a  theoretically and
observationally motivated range exists around  ${\alpha}_{B}$ = -1.6 $\pm 0.2$, thereby
 leading to a ballpark agreement between our estimates and observationally inferred field magnitudes of magnetic clouds (MCs) at L1.}
{In a statistical sense, our method  provides results
that are consistent with observations.}
\date{Received ........ / Accepted .......}

\authorrunning{Patsourakos \& Georgoulis}
\titlerunning {Parametric study to infer near-Sun CME and 1 AU ICME magnetic fields}
\maketitle

\section{Introduction}
\label{sec:intro}

Knowledge of the magnetic field entrained in coronal mass ejections (CMEs) 
is a crucial parameter
for their energetics, dynamics, structuring, and eventually of their
geoeffectiveness. For instance, the  overall CME energy
budget is dominated by the energy stored in non-potential magnetic fields 
\citep[e.g.,][]{forbes2000,avour2000}. In addition, given that
CMEs and interplanetary (IP) counterparts (interplanetary CMEs (ICMEs)) 
are magnetic configurations with a low-$\beta$ plasma parameter, their structural evolution as they propagate and  expand into the IP space is dictated
by the balance and  interactions between their magnetic field
and the ambient solar wind \citep[e.g.,][]{dem2009}.
Moreover, upon arrival at 1 AU, the magnitude of the southward magnetic
field of earth-directed interplanetary CMEs (ICMEs)  is the most important parameter determining 
their geoeffectiveness \citep[e.g.,][]{wu2005}.
Therefore, the near-Sun magnetic field magnitude is a key parameter
for both space weather studies and applications, for example, by constraining
the properties of coronal flux ropes  ejected into the IP medium \citep[e.g.,][]{shiota2016}, 
and in anticipation of the observations of upcoming solar and heliospheric
missions.
Unfortunately,  very few direct observational inferences
of near-Sun  
($\sim$ 1-7 \rs)
CME magnetic fields exist currently  \citep[e.g.,][]{bast2001,jensen2008,tun2013}. These are based on
relatively rare  radio emission configurations, such as  
gyrosynchrotron emission from CME cores and Faraday rotation, and 
require detailed physical modeling of relevant radio emission processes to infer the  magnetic field that is sought after.

We recently proposed a new method to deduce the near-Sun magnetic
field magnitude (hereafter, magnetic field) 
of CMEs 
\citep[][]{case16}. 
This method relies on the conservation of magnetic helicity in cylindrical flux ropes and uses as inputs the magnetic helicity
budget of the source region and geometrical parameters 
(length and radius)
of the associated
CME. It supplies an estimation of the near-Sun CME magnetic field
which is then extrapolated to 1 AU using a power-law fall-off dictated by the radial (heliocentric) distance.
We have  successfully applied this method to a major geoeffective CME 
launched from the Sun on 7 March 2012, which triggered one of the most
intense geomagnetic storms of solar cycle 24.
Recently, two other methods to infer the CME-ICME
magnetic field vectors were proposed \citep[][]{kunkel2010,savani2015}.
\citet{kunkel2010} use a flux-rope CME model, driven by poloidal magnetic flux injection, which 
is constrained by the height-time profile of the associated CME.  The \citet{savani2015} method is  based on 
the heliospheric magnetic helicity rule, the tilt of the source active region (AR), and the magnetic field strength of the compression
region around the CME.

In this work we perform a parametric study of the method to assess its robustness before applying it to observed cases.
We essentially use distributions of input parameters
derived from observations to 
determine the near-Sun and 1 AU magnetic fields for 
a set of synthetic CMEs. This study offers statistics sufficient to 
determine the range of the anticipated
CME magnetic fields both near-Sun and at 1 AU. The latter distribution is compared to actual magnetic-cloud (MC) observations at 1 AU.

In the following, Section \ref{sec:method} describes how we infer the near-Sun CME magnetic field, while Section \ref{sec:b1au} describes how this 
value
is extrapolated to 1 AU. Section \ref{sec:parametric} describes our parametric study, 
Section 5 includes some further tests and uncertainty estimations, 
while Section 6 summarizes our results, their limitations, and an outlook for future revisions.

\section{The helicity-based method to infer the near-Sun magnetic field of CMEs}
\label{sec:method}
\subsection{Theory}
\label{sec:theory}
We  use the Lundquist
flux-rope model \citep{lund1950} as a typical IP prescription of propagating MCs. 
This is an axisymmetric force-free solution with components expressed in cylindrical coordinates ($r,\phi,z$) as
\begin{equation}
B_{r}=0, \, B_{\phi}=\sigma_{H}B_{0}J_{1}(\alpha r),\, B_{z}=B_{0}J_{0}(\alpha r),
\label{eq:lund}
\end{equation}  
where $J_{0}$ and $J_{1}$ are the Bessel functions of the zeroth  and  first
kind, respectively,  ${\sigma}_{H}=\pm 1$  is the helicity sign (i.e, handedness), $\alpha$ is the (constant) force-free parameter, and $B_{0}$ is the maximum (axial) magnetic field. The standard assumption that the first zero
of $J_{0}$ occurs at the edge of the flux rope 
\citep[e.g.,][]{lepping1990}
is made here, namely
\begin{equation} 
\alpha R = 2.405,
\label{eq:alpha}
\end{equation}
with  $R$ corresponding to the flux-rope
radius. This assumption leads to a purely axial or azimuthal magnetic field
at the flux-rope axis or edge.

Following  Equation 9 of  \citet{dasso2006}, the 
magnetic helicity $H_{m}$ of a Lundquist flux rope is written as
\begin{equation}
H_{m}=\frac{4\pi{B_{0}}^{2}L}{\alpha}\int_{0}^{R}{J_{1}}^{2}(\alpha r)dr,
\label{eq:hm}
\end{equation}
where $L$ is the flux-rope length.
The CME magnetic 
field distribution at 15 \rs from an 2.5D MHD simulation
was found to be in excellent agreement with 
the Lundquist model described above \citep[see Figure 8 in][]{lynch2004}.

Solution of  the above equation for the unknown
axial magnetic field $B_{0}$, with the aid of Equation \ref{eq:alpha}, gives
\begin{equation}
B_{0}=\sqrt{\frac{2.405 H_{m}}{4\pi L R J } },   
\label{eq:b1}
\end{equation}
with 
\begin{equation} 
J =\int_{0}^{R}{J_{1}}^{2}(\alpha r)dr.
\label{eq:j}
\end{equation}

Hence, the parameters determining $B_0$, via the application
of Equations \ref{eq:alpha}, \ref{eq:b1}, and \ref{eq:j}, in this case are the length $L$ and radius $R$ of the flux-rope CME along with its magnetic helicity content, $H_m$.

\subsection{Observational constraints to determine the near-Sun CME magnetic field magnitude}
\label{sec:obsconstr}
From the analysis of the previous section, one needs to know a set of magnetic and geometrical properties of a CME to calculate the  axial magnetic field $B_0$.
 In this section we discuss how to deduce estimates of  these parameters from observations.

To infer the magnetic helicity content $H_m$ of a CME, one needs to first calculate the coronal helicity content of the solar source region. This is achieved in various ways. These methods typically use photospheric, mainly vector, magnetograms
and are based on various  theoretical setups, including the calculation of  the 
magnetic helicity-injection rate from photospheric motions \citep[][]{pariat2006},  
partitioning of the photospheric flux into assumed slender flux tubes,
 calculation of the connectivity matrix to deduce the total helicity 
\citep[][]{geor2012}, and classical volume calculations on coronal magnetic field extrapolations 
\citep[][among others]{regcan06,val_etal12,morait2014}. Detailed descriptions of the different methods can be found in the above works.

To obtain the geometrical parameters $R$ and $L$ we 
use the graduated cylindrical shell (GCS) forward fitting model of \citet{thern2009}.
This is a geometrical flux-rope model routinely used
to fit the large-scale appearance 
of flux-rope CMEs in multi-viewpoint 
observations acquired by  the coronagraphs on board the Solar and Heliospheric
Observatory (SOHO) and Solar Terrestrial Relations Observatory (STEREO) spacecraft.
The GCS user modifies a set of free geometrical (front height $H$, 
half-angular width $w$,
aspect ratio $k$, and tilt angle) and positional  (central longitude and latitude) 
parameters
of the flux-rope CME until  a satisfactory agreement is achieved 
between the model projections  and the actual observations. A detailed description can be found in \citet{thern2009}.

In the framework of the GCS model, the CME radius $R$ at a heliocentric distance $r$ is 
\begin{equation}
R(r)=k r.
\label{eq:cmer}
\end{equation}
To assess the flux-rope length $L$, it is assumed  that the CME front 
is  a cylindrical section (see Figure 1 of \citet{dem2009}) with an angular width provided by the geometrical fitting.  
One may then write   
\begin{equation}
L=2 w r_{mid},
\label{eq:cmelength}
\end{equation}
where $r_{mid}(=H-R)$ is the heliocentric distance halfway through the 
model's cross section, along its axis of symmetry. 
The half-angular width $w$ is given in radians.

It is important to realize that the source-region determinations of magnetic helicity, including estimates of the CME helicity content $H_m$, correspond to the 
photosphere or low corona, while those for $R$ and $L$ refer to
the outer corona, which are typically a few solar radii in heliocentric distance. To allow the use of this $H_m$ we adhere to the well-documented conservation principle of magnetic helicity \citep[][]{berger1984, berger1999}. Indeed, for a magnetized plasma with a high magnetic Reynolds number, as the solar corona is widely believed to be, the relative magnetic helicity is conserved even in case of magnetic reconnection; for a recent, successful test of the conservation principle, see \citet{pariat_etal15}. Assuming that an 
ascending CME in the solar corona does not accumulate substantial overlying magnetic structures 
that drastically modify its magnetic helicity content, we use its estimated low-coronal $H_m$ up to the outer corona. 

Summarizing, estimates of $R$, $L$, and $H_m$ allow us to estimate an upper limit of the near-Sun axial magnetic field $B_0$ of flux-rope CMEs at distances covered by coronagraphs.
\section{Extrapolation of the near-Sun CME magnetic field magnitude to 1 AU} 
\label{sec:b1au}

To extrapolate the near-Sun CME magnetic-field magnitude  $B_{*}$, determined at a heliocentric distance $r_{*}$ (Section 2), to 1 AU, 
we assume that its radial evolution follows a power-law behavior of the form  
\begin{equation}
B_{0}(r)=B_{*} {(r/r_{*})}^{{\alpha}_{B}},
\label{eq:scaleb}
\end{equation}
with $r$ corresponding to the heliocentric distance. In Equation \ref{eq:scaleb} we assume that the power-law index ${\alpha}_{B}$ varies in the range [-2.7, -1.0]. This is a typical approximation that is 
frequently followed in the literature \citep[e.g.,][]{patzold1987,kumar96,volker98,vra2004,liu05,forsh06,leitn07,dem2009,poomv2012,mancuso2013,winslow2015,good2016}. 
These theoretical and observational studies also roughly determine the range of ${\alpha}_{B}$ values used here. Most of these studies do not fully cover the  range   (i.e., [10 \rs, 1 AU]) we are considering here,  but typically subsets thereof, either near-Sun or inner heliospheric.

\section{Parametric study}
\label{sec:parametric}
The parameterization of our method consists of the following steps: 
\begin{enumerate}
\item We randomly select a  magnetic helicity value $H_{m}$ resulting 
from a distribution of 162 active-region helicity values 
at different times, corresponding to 42 different solar ARs \citep[][]{tzio2012}. 
The selected active-region helicity is then assigned to a synthetic CME, therefore assuming for simplicity,
that the CME is fully extracting its  source region helicity.
Given the ample dynamical range of the helicity values in the above study (at least three orders of magnitude), even assigning a fraction of each active-region helicity value to model the CME helicity would not lead to remarkably different statistical results. 
\item We randomly select  CME aspect ratios and  angular widths from distributions  resulting from the  forward modeling  of  65  CMEs 
observed by the STEREO coronagraphs \citep[][]{thern2009,bosman2012}. The
observations correspond to a distance of 10 \rs, therefore supplying near-Sun
geometric properties of the observed CMEs.
The GCS model of \citet{thern2009}, described in Section 2, was used in the analysis of these observations.
The deduced CME aspect ratios and angular widths take values in the intervals [0.09,0.7] and [6,41] 
degrees, respectively.
We then deduce the corresponding radii $R$ and lengths $L$ from Equations  \ref{eq:cmer} and \ref{eq:cmelength}, respectively.
\item From the above information, we calculate a near-Sun CME magnetic field $B_{*}$ at $r_{*}=10$ \rs$\;$(Equation \ref{eq:b1}). 
\item For the $B_{*}$ calculated in step 3,  
and for each  of the 18 equidistant values with a step equal to 0.1 covering
the   ${\alpha}_{B}$ range ([-2.7,-1.0]) described in the previous section, 
we determine  18  CME magnetic field ($B_{1AU}$) values at $r=1 AU$ from Equation \ref{eq:scaleb}.
\item We repeated ${10}^{4}$ times the process of randomly selecting  
$Hm$, $R,$ and $L$ to get a corresponding $B_{*}$ (steps 1-3). 
This supplied sufficient statistics to build a  database of synthetic CMEs.
The combination
of the  ${10}^{4}$ near-Sun CME magnetic field values with the 18 different ${\alpha}_{B}$ values
gave rise to 180,000  total values of the CME magnetic field at 1 AU. 
\end{enumerate}
In essence, the above  parameterization provides $10^4$ near-Sun CME magnetic fields $B_*$ and, out of those, $1.8 \times 10^5$ ICME magnetic fields $B_{1AU}$ at 1 AU.

\begin{figure}[!h]
\begin{center}
\includegraphics[width=0.4\textwidth]{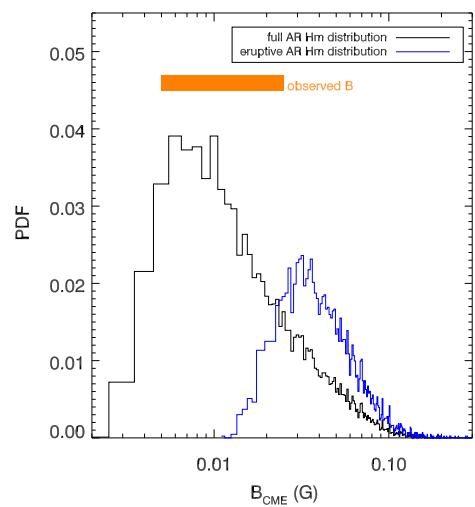}
\end{center}
 \caption{Probability density functions 
of the derived near-Sun CME magnetic fields for $10^4$ synthetic CMEs in two different cases: 
using the sample of all (eruptive and non-eruptive) active-region relative magnetic helicity budgets $H_m$ (black histogram) and using only the subsample of eruptive active-region helicity budgets (blue histogram). The horizontal orange bar shows the range of various observational estimates for the magnetic field of the quiescent (i.e., noneruptive) solar corona. In all cases, the estimates correspond to a heliocentric distance of 10 \rs.
} 
 \label{fig:near_sun}
\end{figure}

\begin{figure}[!h]
\includegraphics[width=0.4\textwidth]{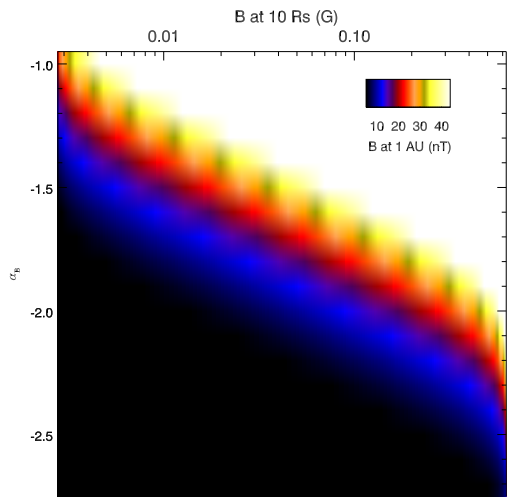}
 \caption{Color-coded range of the derived $B_{1AU}$ (nT) as a function
of the near-Sun CME magnetic field $B_0$ at  10 $ R_{\odot}$ (abscissa) and of the power-law exponent ${\alpha}_{B}$ of the radial CME-ICME falloff (ordinate). The color scale is saturated such that white and black areas lie outside the observed MC magnetic fields by WIND observations at L1.
}
\label{fig:b1aucolor}
\end{figure}

In Figure \ref{fig:near_sun} we show the probability density function (PDF) of the derived near-Sun CME magnetic fields $B_{*}$ at $r_{*}= 10$ \rs$\;$in two different situations: using all active-region helicity values of \citet[][]{tzio2012}  (black histogram) and using only the active-region helicity values of eruptive regions 
(i.e., hosting flares of GOES class M1.0 and above), in which case these values exceed $2\times{10}^{42} \mathrm{{Mx}^{2}}$ (blue histogram). 
In the first case, the PDF peaks at $\approx$ 0.007 G and has a full width at half maximum (FWHM) range at roughly [0.004, 0.03] G. The distribution is asymmetric, showing an extended $B_{*}$  tail. In the second case, the PDF peaks at higher values, $\sim$ 0.03 G, and 
presents a FWHM at roughly [0.02, 0.06] G. 
This distribution corresponds to ARs that are known to be more prone to eruptions  
\citep[e.g.,][]{andrews2003,nindos2015}.

Weaker flares do not necessarily mean lower helicity budgets as an eruptive AR with substantial magnetic helicity can give a series of eruptive C-class flares along with M- and, possibly, X-class flares.

\begin{figure}[!h]
\begin{center}
\includegraphics[width=0.4\textwidth]{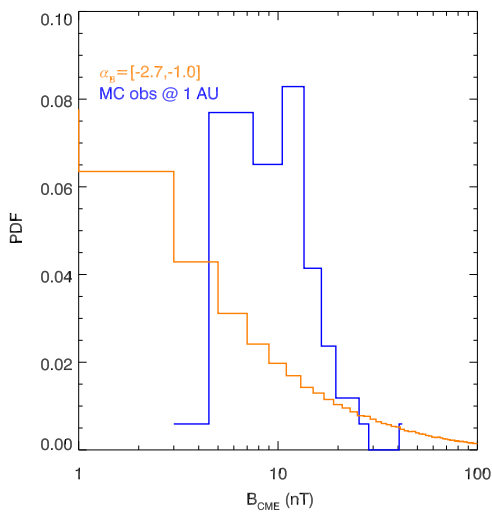}
\end{center} 
\caption{Probability density function of the extrapolated to 1 AU magnetic fields 
of 10,000 synthetic CMEs (orange histogram). These functions correspond to the full range 
of the  considered ${\alpha}_{B}$ values, i.e., 180,000 $B_{1 AU}$ values in total. 
This is compared with the probability density function 
of the magnetic-field magnitude for 162 MCs observed in situ at 1 L1 by WIND (blue histogram).
} 
 \label{fig:b1auall}
\end{figure}

\begin{figure}[!h]
\begin{center}
\includegraphics[width=0.4\textwidth]{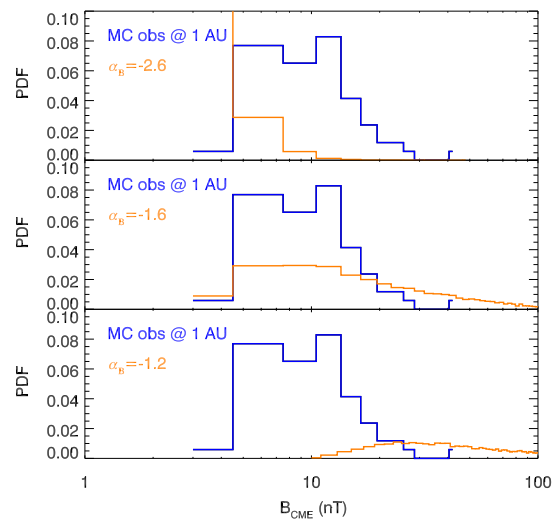}
\end{center}
 \caption{Probability density functions of the extrapolated to 1 AU magnetic field for 10,000 
synthetic CMEs (orange histogram). 
The probability density functions correspond to ${\alpha}_{B}$ equal to -2.6 (top plot), -1.6 (middle plot), and -1.2 (bottom plot). All cases are compared with the PDF of the magnetic field magnitude for 162 magnetic observed in situ at  L1 by WIND (blue histogram).}
\label{fig:b1au0}
\end{figure}

There are a few observational inferences
of the coronal magnetic field at  10 $R_{\odot}$. They rely on  techniques such as Faraday rotation and 
CME-shock stand-off distance
and give magnetic field strengths in the range [0.009-0.02] G 
\citep[e.g.,][]{bempo2010,gopal11,kim2012,poomv2012,mancuso2013,susino2015}.  
They  mainly correspond to observations in the quiescent corona and  
are represented by the orange horizontal bar in Figure \ref{fig:near_sun}. 
A significant fraction of the synthetic CMEs  have
magnetic fields comparable to  or higher than those corresponding 
to the quiescent corona. The latter is essentially the case
for the subset of synthetic CMEs that correspond to prone-to-erupt
ARs. Our results are thus consistent with the notion that CMEs
are structures with stronger magnetic fields than the quiescent ambient corona.

A context representation of the extrapolated CME-ICME magnetic fields at 
1 AU, $B_{1 AU}$, for the 180,000 considered cases, is given in Figure \ref{fig:b1aucolor}. Here we use a color representation of 
$B_{1 AU}$ as a function of $B_{*}$ and ${\alpha}_{B}$.
The color scaling has been saturated so that $B_{*}$ values outside
the range of magnetic field magnitudes in 
observed magnetic clouds (MCs) at 1 AU, namely $B_{MC} \in [4,45]$ nT, 
are shown in either black (smaller) or white (higher). 
Any other color corresponds to projected $B_{*}$ values within the observed $B_{MC} $ range. The distribution of $B_{MC}$ results from linear force-free fits of 162 MCs observed in situ at 1 AU by WIND
\citep[][]{lynch2003,lepping2006}.
Several remarks can be made from this image. First, there is a significant number of cases, i.e.,
$(B_{*}$- ${\alpha}_{B})$ pairs, resulting in  $B_{1 AU}$ values outside the observed $B_{MC}$
range. 
This suggests that the corresponding parameter space can be significantly constrained. 
Second, $B_{1 AU}$ seems to depend more sensitively on ${\alpha}_{B}$ than on 
$B_{*}$. This can be assessed from Figure \ref{fig:b1aucolor} by noting 
that while the vertical colored (i.e., not black and white) bands  corresponding to a given $B_{*}$ in agreement with the observed $B_{MC}$ range show more or less the same extent, this is not the case for the horizontal colored bands corresponding to a given ${\alpha}_{B}$. In this latter case, we also notice very narrow bands at both ends of the employed ${\alpha}_{B}$ range.

To  better understand  the $B_{1 AU}$ sensitivity on ${\alpha}_{B}$
we perform the following further tests. In Figure \ref{fig:b1auall} we show
the histogram of $B_{1AU}$ (orange curve) corresponding to the full range of the 
considered ${\alpha}_{B}$ values, which is overplotted on the histogram of $B_{MC}$ (blue curve). It is then clear that the two histograms do not match, even saliently. This suggests that not all employed  ${\alpha}_{B}$ values  
yield consistent results, which is in line with the result of Figure \ref{fig:b1aucolor}.

\begin{figure}[!h]
\includegraphics[width=0.4\textwidth]{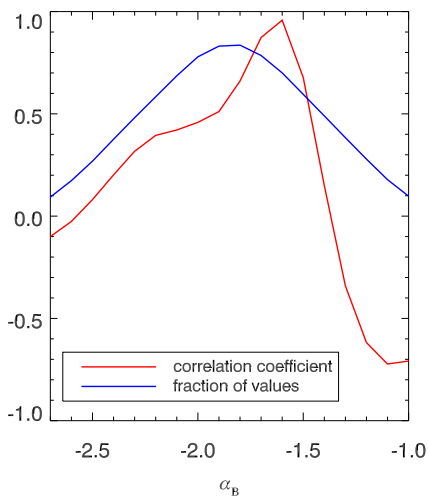}
 \caption{Correlation coefficient of the probability density functions for the predicted $B_{1AU}$ and observed $B_{MC}$ values at L1 as a function of ${\alpha}_{B}$ (red curve). Also shown is the respective fraction of ${B_{1 AU}}$ values (blue curve) falling within the
observed $B_{MC}$ range.}
 \label{fig:bcc}
\end{figure}

In  Figure \ref{fig:b1au0} we show the $B_{1 AU}$
histograms corresponding to three different, specific values of ${\alpha}_{B}$.
These values were meant to represent the two extremes
of the ${\alpha}_{B}$ distribution, but also a value maximizing the reproduction of $B_{MC}$ by $B_{1AU}$.
Radial falloffs of the CME-ICME magnetic field that are too steep (${\alpha}_{B}$=-2.6; top plot)
or too shallow (${\alpha}_{B}$=-1.2; bottom plot) give rise to $B_{1AU}$ values that are too low and
too high, respectively, compared with the MC observations.
On the other hand, setting  ${\alpha}_{B}$=-1.6 (middle plot) 
we obtain a fair agreement between the predicted and observed
CME magnetic fields at 1 AU, at least for the bulk of the distribution.  Both distributions peak
around 10 nT and have similar FWHM of $\sim$ 15 nT. However, the modeled distribution has a high-B tail that is
not present in the MC observations.
A similar value for
${\alpha}_{B}$ was found in an application of the method
to a single event \citep{case16}. 

Clearly, there is a range of ${\alpha}_{B}$ values around -1.6 that yields results  that are consistent with MC observations. We produced Figure \ref{fig:bcc} to firmly establish this interval and quantify its merit. In this test we show the linear correlation coefficient (red curve) between $B_{MC}$ and $B_{1U}$ histograms as a function of ${\alpha}_{B}$, thus obtaining 18 values of this correlation coefficient. We find that the correlation coefficient  exhibits a well-defined peak around 0.9 at ${\alpha}_{B}$=-1.6 
and stays above 0.5 when ${\alpha}_{B} \in$ [-1.9,-1.5]. 
Very small or even negative correlation coefficients are found when ${\alpha}_{B}$ moves toward extreme values of its assumed range.

Another useful measure of best-fit ${\alpha}_{B}$ values is provided by the fraction of the projected or predicted values $B_{1AU}$ agreeing with the range of $B_{MC}$ values as a function of ${\alpha}_{B}$. This is shown by the blue curve in Figure \ref{fig:bcc}. The peak of this fraction occurs at 0.8 (80 \%) for a slightly different ${\alpha}_{B}$ value (-1.9) compared to the peak of the correlation coefficient (-1.6). Nonetheless, the fraction is above 0.5 (50 \%) for ${\alpha}_{B} \in$ [-1.9,-1.5]. The relative discrepancy between the peaks of the two curves in Figure \ref{fig:bcc} is not unexpected. Indeed, the correlation coefficient measures the degree of overlap between the two distributions, while the fraction denotes the subset of points within a given range with no a priori reason for the two distributions to match. Since both the fraction and correlation coefficient reach their maxima for $\alpha_{B} \in$ [-1.9,-1.5],  however, we consider this range as the best-fit range, as we are  able to reproduce the observed $B_{MC}$ distribution  relatively well  and at the same time yield a significant number of cases within the $B_{MC}$ range.

\section{Further tests and an uncertainty estimation for $\alpha _B$}
\label{sec:tests}

An important issue, which is directly relevant to our analysis, is how the (input) AR helicity 
PDF relates to the MC helicity content at 1 AU. To investigate this, we constructed the 
PDF of the magnetic helicity of MCs observed at 1 AU (see also Lynch et al. (2005) and 
\citet{dem2016} for MC $H_m$ PDFs) and compared it with the AR helicities 
used in this study. We used the Lundquist linear force-free model, as 
in our analysis, to obtain MC helicities and applied this model to the MC fittings of Lynch et al. (2003) and Lepping
et al. (2006). 
We used two different approaches for the MC lengths required in the Hm calculation. First,  we used the results in situ observations of near-relativistic
electrons inside MCs, which can supply a proxy for their lengths, given their solar release times, onset
times at 1 AU, and speeds (e.g., Larson et al. 1997). This is because near-relativistic
electrons, assuming they propagate scatter-free, have small gyroradii and thus follow the magnetic field very
closely. A statistical study of 30 near-relativistic events in 8 MCs gave 
an average MC length of 2.28 AU (Kahler et al. 2011).  Second, we adopted the statistical approach of \citet{dem2016}. 
This study used the results of MC fittings of Lynch et al. (2003)
and Lepping et al. (2006) and found that several MC properties, including their helicity per
unit length, do not depend significantly on the position along the MC axis. This allowed the 
derivation of a generic shape for the MC axis, parameterized in terms of its angular span, which 
further enabled an estimation of the MC length assuming a rooting of both its legs in the Sun. 
The resulting average MC length was 2.6 AU.  The average of the two MC-length estimates  above, 
which we use for  the remainder of this Section, is 2.44 AU.

\begin{figure}[!h]
\begin{center}
\includegraphics[width=0.4\textwidth]{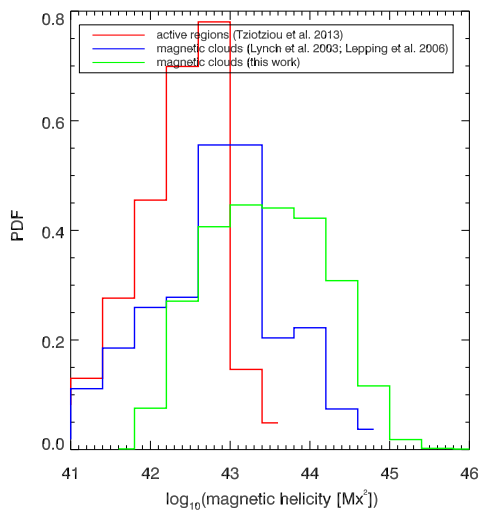}
\end{center}
 \caption{Probability density functions of the Hm corresponding to (a) AR sample of Tziotziou et al. (2013) (red histogram), (b)  MC linear-force
 free fittings at 1 AU using the data from the Lynch et al. (2003) and Lepping et al. (2006) studies (blue histogram), and (c) 
 $B_{1AU}$ corresponding to the best-fit ${\alpha}_{B}$ from this work (green histogram).
} 
 \label{fig:helicity}
\end{figure}

We further constructed a helicity PDF for the synthetic MCs of this study. For this task we used the 
$B_{1AU}$ values of these MCs, an average MC length of 2.44 AU, and an average MC radius of 
0.11 AU, as obtained by  Lynch et al. (2003) and Lepping et al. (2006). Figure 6 depicts the 
resulting MC $H_m$ distribution from (i) the above literature studies (blue histogram), 
(ii) the source ARs (red histogram), and (iii) our synthetic MCs (green histogram). 
The maximum-likelihood values  and FWHMs (in ${10}^{42} \mathrm{{Mx}^{2}}$)  of the three distributions are 
6.3  
and 5.3
for the AR
$H_m$ distribution; 6.3 and 13.3 for the literature work estimates; 
and 15.8 and 248.7 for the synthetic MC $H_m$ of this study.

Clearly, our AR  $H_m$ distribution shows a deficit compared to both observed and synthetic MC 
$H_m$ distributions. With respect to the observed MC distribution, this is 
not totally unexpected, as both AR and MC $H_m$ calculations are model dependent, 
and possibly involve various systematic effects. For example, Tziotziou et al. (2013) calculated AR 
helicities using the Georgoulis et al. (2012) method that, by construction, infers a lower limit of AR 
free energies and the corresponding relative magnetic helicity. Different helicity calculation methods in 
ARs give rise to difference factors ranging between 1 and several (but less than 10) units. An 
analysis by Tziotziou et al. (2016, in preparation), in particular, gave a difference factor of $\sim 2.5$, while 
Tziotziou et al. (2013) and Nindos \& Andrews (2004) independently found average AR helicities on 
the order 6.6 $\times 10^{42}$ $\mathrm{Mx}^2$ and 19.5 $\times 10^{42}$ $\mathrm{Mx}^2$,  which also differ by a factor of $\sim$3.
In addition, one cannot exclude the possibility that CMEs accumulate more helicity during their initial stages 
in the inner corona by poloidal magnetic flux addition via magnetic reconnection with their surroundings 
(e.g., Lin et al. 2004; Qiu et al. 2007) and, conversely, loose helicity in the inner heliosphere owing to 
magnetic erosion (e.g., Dasso et al. 2006; Gosling et al. 2007; Manchester et al. 2014; 
Ruffenach et al. 2015). Moreover, the flux-rope structure (i.e., twisted
magnetic fields) may be confined only to the MC leading edge, (e.g., Owens 2016), suggesting that
the employed magnetic field lengths in MC $H_m$ calculations could represent upper limits. Finally, 
while different cylindrical
MC models applied to the same data set lead to rather small differences 
in the resulting helicities 
(up to $\sim$  30 $\%$ ; Gulisano et al. (2005)), 
departures from circular MC cross sections could lead to 
larger (by a factor 2-3) differences \citep{dem2016}.

Comparing the synthetic and observed MC $H_m$ distributions from Figure 6, we see than the
former corresponds to somewhat higher values compared to the latter. Given the significant overlap,
however, we could say that the distributions are not dissimilar. Again, this is not totally unexpected
because of systematic effects involved in calculations. In addition, these differences may be due to the fact that our method does not explicitly invoke the near-Sun-1 AU helicity conservation, however, a power-law
formulation that has been introduced in line with many previous studies, 
but in a rather ad hoc manner.

Concluding, it seems reasonable to expect some differences in the statistical distributions of the 
AR and MC helicities. That said, several studies found an overall agreement, albeit with 
significant uncertainties, between the source region eruption-related and associated MC helicities
(e.g., Green et al. 2002; Nindos et al. 2003; Luoni et al. 2005; Mandrini et al. 2005; Rodriguez
et al. 2008; Kazachenko et al. 2012). This rough $H_m$ conservation was found not only between
the Sun and 1 AU, but also for a MC observed at 1 AU and at 5.4 AU (Nakwacki et al. 2011).
Finally, statistical studies found that the $H_m$ signs of the CME source regions matches those
of associated MCs for up to 88 $\%$ (Bothmer $\&$  Schwenn 1998; Cho et al. 2013). These studies
underline the connection between the source region and MC Hm, at the same time shedding light on the 
significant uncertainties present in all stages of the calculations. This remains an objective for future 
efforts to narrow down and constrain the various uncertainties and modulations.

Further on, we briefly investigate the sensitivity of the near-Sun $B_0$ value to the input 
AR helicity. This was achieved by overestimating and underestimating the AR $H_m$ values 
by factors of 3 and 1/3, for reasons explained above. Factors 3 (1/3) give rise to higher (lower) 
near-Sun magnetic fields, and therefore steeper (shallower) radial falloffs of the CME 
magnetic field in the IP space are required in order to match the
observed magnetic-field range in MCs. The values of the power-law index $\alpha _B$ yielding a 
maximum correlation between the predicted and observed MC magnetic field are -1.8 and -1.4, 
respectively. These $\alpha _B$ values correspond to rather small departures from the best-fit 
$\alpha _B$ of -1.6 and could thus serve as a measure of the uncertainty ($\pm 0.2$) of
 the best-fit $\alpha _B$.

\section{Summary and discussion}
\label{sec:discussion}
Developing methods for the practical estimation of the 
magnetic field of CMEs, both near the Sun and at 1 AU,
is a timely and important task for assessing the near-Sun energetics and dynamics of CMEs and for providing clues of the possible geoeffectiveness of their ICME counterparts. We recently developed one such method and 
we hereby perform a parametric study of it. 
Our study only applies to the CME magnetic-field magnitude and not its orientation, hence, 
reaching results pertinent to the CME geoeffectiveness requires an extension of this work.
Our major conclusions are the following:
\begin{enumerate}
\item The predicted near-Sun CME (at 10 \rs) magnetic fields (Figure \ref{fig:near_sun}) exhibit a FWHM range of [0.004, 0.03] G
and their distribution  shows values that are comparable to, or higher
than, magnetic fields measured in the quiescent corona by a handful of observations. For solar AR 
sources prone to eruptions, the FWHM of CME magnetic fields is [0.02, 0.07] G, which is clearly higher than the quiescent-corona magnetic field at 10 \rs$\;$(Figure \ref{fig:near_sun}).  
\item The extrapolated CME-ICME magnetic field at 1 AU depends more sensitively on the power-law index ${\alpha}_{B}$ of its radial dependence than on the near-Sun CME magnetic fields (Figure \ref{fig:b1aucolor}).
\item Considering the full range of literature-suggested ${\alpha}_{B}$ values  ([-2.7,-1.0]), we find that the extrapolated near-Sun magnetic fields at 1 AU 
do not match MC magnetic-field measurements  (Figure \ref{fig:b1auall}).
\item For $\alpha_{B}$ varying in the range [-1.9,-1.4], we obtain a considerable ballpark agreement with MC magnetic-field measurements at 1 AU in terms of both the similarity of the corresponding distributions and the high fraction of $B_{1AU}$ values falling within the $B_{MC}$ value range. A best-fit $\alpha_{B}$ attains a value of -1.6 (Figures \ref{fig:b1au0}, \ref{fig:bcc}). 
\end{enumerate}

Statistically, therefore, our method is able to reproduce the ballpark of the ICME magnetic field magnitudes at 1 AU reasonably
well. 
This result encourages us to seek further opportunities to  apply the method to observed CME cases in the future. 
Interestingly, the best-fit $\alpha_{B}=$-1.6 stems independently from the analytical 
model of \citet{dem2009}, which treats CMEs as expanding force-free magnetic flux ropes 
in equilibrium with the total pressure of the ambient solar wind.

In the following, we summarize our assumptions and simplifications that could represent areas of future method improvements. 
We used AR helicity values taken from \citet{tzio2012}, which were calculated via the \citet{geor2012} method. Several methods exist to calculate $H_{m}$ (Section \ref{sec:theory}). Application of these methods to the same data set, i.e., a sequence of HMI vector magnetograms, 
spanning over a two-day period (6 - 7) of March 2012 for the supereruptive NOAA AR 11429, showed that the $H_{m}$ determinations, even thought they stem from very different methods, show an overall
agreement within a factor $\sim 2.5$ (Tziotziou et al., 2016, in preparation). 
In addition, while the employed $H_{m}$ values refer to entire ARs, it is known that no AR sheds 
its entire helicity budget in a single eruption; it is more appropriate to attribute a fraction of the helicity budget  to eruptions. This fraction seems to be relatively small, 
typically one order of magnitude less, but can be up to 40$\%$ of
the total helicity budget in some models 
\citep[e.g.,][]{kliem2011,morait2014}. Nevertheless, as a first-order approximation, the used AR-wide $H_{m}$ value should not dramatically overestimate the CME magnetic field in our analysis, given the large dispersion of helicity values ($\sim$3 orders of magnitude) and the large number of synthetic CMEs ($10^4$). In future works, nonetheless, it is meaningful to search for eruption-related $H_{m}$ changes that should then be attributed to the ensuing CME.

As observed in coronagraph field of view, CMEs have curved fronts. In Section \ref{sec:theory} we nonetheless assumed a straight, cylindrically shaped CME front. This is  because  the employed flux-rope model is cylindrical. However, CMEs may flatten during IP propagation  \citep[e.g.,][]{savani2010}. At any rate, adopting a curved CME-front shape most likely introduces a (small) scaling factor
in the derived $B_{*}$ distributions.

In the  parametric model 
of Section \ref{sec:parametric}, we
assumed that the distributions of the magnetic ($H_{m}$)
and geometrical parameters ($\alpha$ and $\kappa$) 
are statistically independent, that is, 
they do not exhibit statistical correlations.
This may be  not entirely true.   

In spite of the agreement found, the FWHM range of ${\alpha}_{B}$ values still allows $\sim$30 \%$\;$ of the projected $B_{1AU}$ values to lie outside the observed $B_{MC}$ value range (Figure \ref{fig:bcc}). For example, there is a high-$B_{1AU}$ tail that is not present in the observations (Figure \ref{fig:b1au0}). This suggests that our single (and simple) power-law
description of  the radial evolution of CME-ICME magnetic field
of Equation \ref{eq:scaleb} may require future improvement. In particular, it is possible that the CME magnetic field experiences a 
stronger radial decay closer to the Sun, hence a single-power law description may not be entirely realistic. 
In addition, CMEs-ICMEs could experience
magnetic erosion during their IP travel  
\citep[e.g.,][]{dasso2006,ruffe2015}, and they may thus end up with reduced magnetic fields at 1 AU. Finally, while excessively high MC magnetic fields are only very rarely reported \citep[e.g.,][]{liu2014}, the lowest  $B_{1AU}$ values, below the lower limit of the
$B_{MC}$ distribution, may call for complex processes in the IP medium 
that could enforce the magnetic field of an ICME. This could be 
a CME-CME interaction, for example, or an interaction between a CME and trailing fast solar wind streams 
\citep[e.g.,][]{lugaz2008,shen2011,harrison2012,temmer2012,liu2014,lugaz2014}.

Clearly, more detailed analysis is required to tackle the above issues. Such a major task would be the application of the method to a set of carefully selected, well-observed CME-ICME cases, where both photospheric coverage would be sufficient and detailed GCS modeling would exist, along with satisfactory MC measurements at L1. This exercise would also aim to attribute eruption-related helicity changes to CMEs, hence tackling issues (1), (2) above. In addition, it would enable one to determine whether magnetic and geometrical parameters are correlated, therefore addressing issue (3) above. Further theoretical and modeling work is required to understand the radial evolution of CME-ICME magnetic fields, hence tackling issue (4) above. One such avenue would be to to analyze simulations of CME propagation
in the IP medium and monitor the evolution
of their magnetic fields with heliocentric distance, and at the same time  investigating whether and how ${\alpha}_{B}$ depends on CME properties (e.g., speed, width), background solar wind (e.g., speed, density), and IP magnetic field. Here we treated ${\alpha}_{B}$ in a rather ad hoc manner; however, ${\alpha}_{B}$ appears as the single 
most important parameter for describing the ICME magnetic
field at 1 AU, which apparently enables one to encapsulate most of the relevant physics into a simple form of self-similar IP expansion. 
That said, one should not dismiss the role
of the near-Sun CME magnetic field in the determination of the ICME
magnetic field at 1 AU. We also need observational inferences of this important parameter 
at sufficient numbers and our proposed method is one such promising avenue. 
Our framework may be generalized  
to non-force-free states \citep[e.g.,][]{hidalgo2002,chen2012,berdi2013,subra2014,case16,nieves2016}
and cylindrical geometries (i.e., curved flux-ropes) \citep[e.g.,][]{janv2013,vand2015} provided
that the existence of explicit relationships connect its geometrical 
($R$ and $L$)
and magnetic ($H_m$) parameters.
Ultimately, we will perform the most meaningful tests of the two 
central parameters, $B_{*}$ and ${\alpha}_{B}$, and key assumptions of our model, when pristine 
observations by the two forthcoming, flagship heliophysics missions, Solar Orbiter and Solar Probe Plus,  become available.

\section*{Acknowledgements}
The authors extend their thanks to the referee  for important comments and suggestions.
This research has been partly co-financed by the European Union (European
Social Fund -ESF) and Greek national funds through the Operational Program
``Education and Lifelong Learning" of the National Strategic Reference
Framework (NSRF) -Research Funding Program: ``Thales. Investing in knowledge
society through the European Social Fund".
SP acknowledges support from an FP7 Marie Curie 
Grant (FP7-PEOPLE-2010-RG/268288). 
MKG wishes to acknowledge support  from the EU's Seventh Framework Programme under grant agreement no PIRG07-GA-2010-268245. 
The authors acknowledge the  Variability of the Sun and Its Terrestrial Impact
(VarSITI) international program.


\begin{thebibliography}{}

\bibitem[Andrews(2003)]{andrews2003} Andrews, M.~D.\ 2003, 
\solphys, 218, 261 


\bibitem[Bastian et al.(2001)]{bast2001} Bastian, T.~S., Pick,
M., Kerdraon, A., Maia, D., \& Vourlidas, A.\ 2001, \apjl, 558, L65

\bibitem[Berger(1984)]{berger1984} Berger, M.~A.\ 1984,
Geophysical and Astrophysical Fluid Dynamics, 30, 79

\bibitem[Berger (1999)]{berger1999} Berger, M. A. \ 1999, Plasma Phys. Contrl. Fusion, 41, B167




\bibitem[Bemporad
\& Mancuso(2010)]{bempo2010} Bemporad, A., \& Mancuso, S.\ 2010, \apj, 720, 130


\bibitem[Berdichevsky(2013)]{berdi2013} Berdichevsky, D.~B.\ 2013, \solphys, 284, 245 

\bibitem[Bosman et al.(2012)]{bosman2012} Bosman, E., Bothmer, V.,
Nistic{\`o}, G., et al.\ 2012, \solphys, 281, 167

\bibitem[Bothmer
\& Schwenn(1998)]{volker98} Bothmer, V., \& Schwenn, R.\ 1998, Annales Geophysicae, 16, 1



\bibitem[Chen(2012)]{chen2012} Chen, J.\ 2012, \apj, 761, 179 

\bibitem[Cho et al.(2013)]{cho2013} Cho, K.-S., Park, S.-H., Marubashi, K., et al.\ 2013, \solphys, 284, 105

\bibitem[Dasso et
al.(2006)]{dasso2006} Dasso, S., Mandrini, C.~H., D{\'e}moulin, P., \& Luoni, M.~L.\ 2006, \aap, 455, 349

\bibitem[D{\'e}moulin 
\& Dasso(2009)]{dem2009} D{\'e}moulin, P., \& Dasso, S.\ 2009, \aap, 498, 551 

\bibitem[D{\'e}moulin et al.(2016)]{dem2016} D{\'e}moulin, P., Janvier, M., \& Dasso, S.\ 2016, \solphys, 291, 531 

\bibitem[Forbes(2000)]{forbes2000} Forbes, T.~G.\ 2000, \jgr, 105, 
23153 



\bibitem[Forsyth et al.(2006)]{forsh06} Forsyth, R.~J.,
Bothmer, V., Cid, C., et al.\ 2006, \ssr, 123, 383




\bibitem[Georgoulis et al.(2012)]{geor2012} Georgoulis, M.~K., 
Tziotziou, K., \& Raouafi, N.-E.\ 2012, \apj, 759, 1 




\bibitem[Gopalswamy 
\& Yashiro(2011)]{gopal11} Gopalswamy, N., \& Yashiro, S.\ 2011, \apjl, 736, L17 



\bibitem[Gosling et al.(2007)]{gosli2007} Gosling, J.~T., Eriksson, S., McComas, D.~J., Phan, T.~D., \& Skoug, R.~M.\ 2007, Journal of Geophysical Research (Space Physics), 112, A08106 



\bibitem[Gulisano et al.(2005)]{guli2005} Gulisano, A.~M., Dasso, S., Mandrini, C.~H., \& D{\'e}moulin, P.\ 2005, Journal of Atmospheric and Solar-Terrestrial Physics, 67, 1761


\bibitem[Green et al.(2002)]{green2002} Green, L.~M., L{\'o}pez fuentes, M.~C., Mandrini, C.~H., et al.\ 2002, \solphys, 208, 43


\bibitem[Good \& Forsyth(2016)]{good2016} Good, S.~W., \& Forsyth, R.~J.\ 2016, \solphys, 291, 239



\bibitem[Harrison et al.(2012)]{harrison2012} Harrison, R.~A., 
Davies, J.~A., M{\"o}stl, C., et al.\ 2012, \apj, 750, 45 




\bibitem[Hidalgo et al.(2002)]{hidalgo2002} Hidalgo, M.~A., Cid, 
C., Vinas, A.~F., 
\& Sequeiros, J.\ 2002, Journal of Geophysical Research (Space Physics), 107, 1002 






\bibitem[Janvier et 
al.(2013)]{janv2013} Janvier, M., D{\'e}moulin, P., \& Dasso, S.\ 2013, \aap, 556, A50 




\bibitem[Jensen 
\& Russell(2008)]{jensen2008} Jensen, E.~A., \& Russell, C.~T.\ 2008, \grl, 35, 2103 





\bibitem[Kahler et al.(2011)]{kahler2011} Kahler, S.~W., Haggerty, D.~K., \& Richardson, I.~G.\ 2011, \apj, 736, 106 

\bibitem[Kazachenko et al.(2012)]{kaza2012} Kazachenko, M.~D., Canfield, R.~C., Longcope, D.~W., \& Qiu, J.\ 2012, \solphys, 277, 165 

\bibitem[Kim et al.(2012)]{kim2012} Kim, R.-S., Gopalswamy, N., 
Moon, Y.-J., Cho, K.-S., \& Yashiro, S.\ 2012, \apj, 746, 118 


\bibitem[Kliem et al.(2011)]{kliem2011} Kliem, B., Rust, S., 
\& Seehafer, N.\ 2011, IAU Symposium, 274, 125 

\bibitem[Kumar
\& Rust(1996)]{kumar96} Kumar, A., \& Rust, D.~M.\ 1996, \jgr, 101, 15667



\bibitem[Kunkel 
\& Chen(2010)]{kunkel2010} Kunkel, V., \& Chen, J.\ 2010, \apjl, 715, L80 



\bibitem[Larson et al.(1997)]{larson1997} Larson, D.~E., Lin, R.~P., McTiernan, J.~M., et al.\ 1997, \grl, 24, 1911 



\bibitem[Leitner et al.(2007)]{leitn07} Leitner, M., Farrugia,
C.~J., M{\"o}Stl, C., et al.\ 2007, Journal of Geophysical Research (Space
Physics), 112, A06113





\bibitem[Lepping et al.(1990)]{lepping1990} Lepping, R.~P., 
Burlaga, L.~F., \& Jones, J.~A.\ 1990, \jgr, 95, 11957 





\bibitem[Lepping et al.(2006)]{lepping2006} Lepping, R.~P., 
Berdichevsky, D.~B., Wu, C.-C., et al.\ 2006, Annales Geophysicae, 24, 215


 
\bibitem[Lin et al.(2004)]{lin2004} Lin, J., Raymond, J.~C., \& van Ballegooijen, A.~A.\ 2004, \apj, 602, 422


\bibitem[Liu et
al.(2005)]{liu05} Liu, Y., Richardson, J.~D., \& Belcher, J.~W.\ 2005, \planss, 53

\bibitem[Liu et al.(2014)]{liu2014} Liu, Y.~D., Luhmann, J.~G., 
Kajdi{\v c}, P., et al.\ 2014, Nature Communications, 5, 3481 


\bibitem[Luoni et al.(2005)]{luoni2005} Luoni, M.~L., Mandrini, C.~H., Dasso, S., van Driel-Gesztelyi, L., \& D{\'e}moulin, P.\ 2005, Journal of Atmospheric and Solar-Terrestrial Physics, 67, 1734 


\bibitem[Lundquist (1950)]{lund1950} Lundquist, S. 1950, Ark. Fys., 2, 361



\bibitem[Lugaz et al.(2008)]{lugaz2008} Lugaz, N., Manchester, 
W.~B., IV, Roussev, I.~I., 
\& Gombosi, T.~I.\ 2008, Journal of Atmospheric and Solar-Terrestrial Physics, 70, 598 

\bibitem[Lugaz et al.(2014)]{lugaz2014} Lugaz, N., Farrugia, 
C.~J., \& Al-Haddad, N.\ 2014, IAU Symposium, 300, 255 


\bibitem[Lynch et al.(2003)]{lynch2003} Lynch, B.~J., Zurbuchen, 
T.~H., Fisk, L.~A., 
\& Antiochos, S.~K.\ 2003, Journal of Geophysical Research (Space Physics), 108, 1239 


\bibitem[Lynch et al.(2004)]{lynch2004} Lynch, B.~J., Antiochos, 
S.~K., MacNeice, P.~J., Zurbuchen, T.~H., 
\& Fisk, L.~A.\ 2004, \apj, 617, 589 


\bibitem[Lynch et al.(2005)]{lynch2005} Lynch, B.~J., Gruesbeck, J.~R., Zurbuchen, T.~H., \& Antiochos, S.~K.\ 2005, Journal of Geophysical Research (Space Physics), 110, A08107



\bibitem[Manchester et al.(2014)]{manche2014} Manchester, W.~B., Kozyra, J.~U., Lepri, S.~T., \& Lavraud, B.\ 2014, Journal of Geophysical Research (Space Physics), 119, 5449 



\bibitem[Mandrini et al.(2005)]{mandrini2005} Mandrini, C.~H., Pohjolainen, S., Dasso, S., et al.\ 2005, \aap, 434, 725


\bibitem[Mancuso
\& Garzelli(2013)]{mancuso2013} Mancuso, S., \& Garzelli, M.~V.\ 2013, \aap, 553, A100



\bibitem[Moraitis et al.(2014)]{morait2014} Moraitis, K., 
Tziotziou, K., Georgoulis, M.~K., 
\& Archontis, V.\ 2014, \solphys, 289, 4453 

\bibitem[Nakwacki et al.(2011)]{naka2011} Nakwacki, M.~S., Dasso, S., D{\'e}moulin, P., Mandrini, C.~H., \& Gulisano, A.~M.\ 2011, \aap, 535, A52



\bibitem[Nieves-Chinchilla et al.(2016)]{nieves2016} Nieves-Chinchilla, T., Linton, M.~G., Hidalgo, M.~A., et al.\ 2016, \apj, 823, 27

\bibitem[Nindos et al.(2003)]{nindos2003} Nindos, A., Zhang, J., 
\& Zhang, H.\ 2003, \apj, 594, 1033 

\bibitem[Nindos \& Andrews(2004)]{nindos2004} Nindos, A., \& Andrews, M.~D.\ 2004, \apjl, 616, L175 


\bibitem[Nindos et al.(2015)]{nindos2015} Nindos, A., Patsourakos, 
S., Vourlidas, A., \& Tagikas, C.\ 2015, \apj, 808, 117 





\bibitem[Owens(2016)]{owens2016} Owens, M.~J.\ 2016, \apj, 818, 197 


\bibitem[Pariat et 
al.(2006)]{pariat2006} Pariat, E., Nindos, A., D{\'e}moulin, P., \& Berger, M.~A.\ 2006, \aap, 452, 623 

\bibitem[Pariat et al. (2015)]{pariat_etal15} Pariat, E., Valori, G., D\'{e}moulin, P., \& Dalmasse, K. \ 2015, \aap, 580, id.A128

\bibitem[Patsourakos et al.(2016)]{case16} Patsourakos, S., Georgoulis, M. K., Vourlidas, A., et al. \ 2016, \apj, 817 id. 14


\bibitem[Patzold et al.(1987)]{patzold1987} Patzold, M., Bird, 
M.~K., Volland, H., et al.\ 1987, \solphys, 109, 91 


\bibitem[Poomvises et al.(2012)]{poomv2012} Poomvises, W., 
Gopalswamy, N., Yashiro, S., Kwon, R.-Y., 
\& Olmedo, O.\ 2012, \apj, 758, 118 


\bibitem[Qiu et al.(2007)]{qiu2007} Qiu, J., Hu, Q., Howard, T.~A., \& Yurchyshyn, V.~B.\ 2007, \apj, 659, 758
 

\bibitem[R\'{e}gnier \& Canfield (2006)]{regcan06} R\'{e}gnier, S., \& Canfield, R. C. \ 2006, \aap, 451, 319


\bibitem[Rodriguez et al.(2008)]{rodri2008} Rodriguez, L., Zhukov, A.~N., Dasso, S., et al.\ 2008, Annales Geophysicae, 26, 213 

\bibitem[Ruffenach et al.(2015)]{ruffe2015} Ruffenach, A., 
Lavraud, B., Farrugia, C.~J., et al.\ 2015, Journal of Geophysical Research 
(Space Physics), 120, 43 



\bibitem[Savani et al.(2010)]{savani2010} Savani, N.~P., Owens, 
M.~J., Rouillard, A.~P., Forsyth, R.~J., 
\& Davies, J.~A.\ 2010, \apjl, 714, L128 



\bibitem[Savani et al.(2015)]{savani2015} Savani, N.~P.,
Vourlidas, A., Szabo, A., et al.\ 2015, Space Weather, 13, 374





\bibitem[Shen et al.(2011)]{shen2011} Shen, F., Feng, X.~S., 
Wang, Y., et al.\ 2011, Journal of Geophysical Research (Space Physics), 
116, A09103 


\bibitem[Shiota \& Kataoka(2016)]{shiota2016} Shiota, D., \& Kataoka, R.\ 2016, Space Weather, 14, 56 



\bibitem[Subramanian et al.(2014)]{subra2014} Subramanian, P., 
Arunbabu, K.~P., Vourlidas, A., \& Mauriya, A.\ 2014, \apj, 790, 125 


\bibitem[Susino et al.(2015)]{susino2015} Susino, R., Bemporad, 
A., \& Mancuso, S.\ 2015, \apj, 812, 119 

\bibitem[Thernisien et al.(2009)]{thern2009} Thernisien, A., 
Vourlidas, A., \& Howard, R.~A.\ 2009, \solphys, 256, 111 




\bibitem[Temmer et al.(2012)]{temmer2012} Temmer, M., Vr{\v s}nak, 
B., Rollett, T., et al.\ 2012, \apj, 749, 57 

\bibitem[Tun 
\& Vourlidas(2013)]{tun2013} Tun, S.~D., \& Vourlidas, A.\ 2013, \apj, 766, 130 


\bibitem[Tziotziou et al.(2012)]{tzio2012} Tziotziou, K., 
Georgoulis, M.~K., \& Raouafi, N.-E.\ 2012, \apjl, 759, L4 


\bibitem[Tziotziou et al.(2013)]{tzio2013} Tziotziou, K., 
Georgoulis, M.~K., \& Liu, Y.\ 2013, \apj, 772, 115 

\bibitem[Valori et al.(2012)]{val_etal12} Valori, G., D\'{e}moulin, P., \& Pariat, E. \ 2012, \solphys, 278, 347

\bibitem[Vandas 
\& Romashets(2015)]{vand2015} Vandas, M., \& Romashets, E.\ 2015, \aap, 580, A123 

\bibitem[Vourlidas et al.(2000)]{avour2000} Vourlidas, A., 
Subramanian, P., Dere, K.~P., \& Howard, R.~A.\ 2000, \apj, 534, 456 

\bibitem[Vr{\v s}nak et 
al.(2004)]{vra2004} Vr{\v s}nak, B., Magdaleni{\'c}, J., \& Zlobec, P.\ 2004, \aap, 413, 753 

\bibitem[Winslow et al.(2015)]{winslow2015} Winslow, R.~M., Lugaz, 
N., Philpott, L.~C., et al.\ 2015, Journal of Geophysical Research (Space 
Physics), 120, 6101 

\bibitem[Wu
\& Lepping(2005)]{wu2005} Wu, C.-C., \& Lepping, R.~P.\ 2005, Journal of Atmospheric and Solar-Terrestrial Physics, 67, 28


\end{thebibliography}
\end{document}